\documentclass[prl,twocolumn,showpacs,floatfix,preprintnumbers,amsmath,amssymb,superscriptaddress]{revtex4}

\usepackage{graphicx}
\usepackage{dcolumn}
\usepackage{bm}
\usepackage{color}
\usepackage{color}
\usepackage[latin1]{inputenc}
\usepackage[urlcolor=blue]{hyperref}
\hypersetup{backref, colorlinks=true, linkcolor=blue, citecolor=blue}

\begin{document}

\title{{Coexistence of Superconductivity and Charge Density Wave in Tantalum Disulfide: Experiment and Theory}}

\author{Y. Kvashnin}
\affiliation{Uppsala University, Department of Physics and Astronomy, Box 516, SE-751 20 Uppsala, Sweden}

\author{D. VanGennep}
\affiliation{Lyman Laboratory of Physics, Harvard University, Cambridge, Massachusetts 02138, USA}

\author{M. Mito}
\affiliation{Graduate School of Engineering, Kyushu Institute of Technology, Fukuoka 804-8550, Japan}

\author{S. A. Medvedev}
\affiliation{Max Planck Institute for Chemical Physics of Solids, D-01187 Dresden, Germany}

\author{R. Thiyagarajan}
\affiliation{Institut für Festkörper- und Materialphysik, Technische Universität Dresden, 01069 Dresden, Germany}

\author{O. Karis}
\affiliation{Uppsala University, Department of Physics and Astronomy, Box 516, SE-751 20 Uppsala, Sweden}

\author{A. N. Vasiliev}
\affiliation{Ural Federal University, Yekaterinburg 620002, Russia}
\affiliation{Lomonosov Moscow State University, Moscow 119991, Russia}

\author{O. Eriksson}
\affiliation{Uppsala University, Department of Physics and Astronomy, Box 516, SE-751 20 Uppsala, Sweden}
\affiliation{School of Science and Technology, Örebro University, SE-701 82 Örebro, Sweden}

\author{M. Abdel-Hafiez}
\affiliation{Uppsala University, Department of Physics and Astronomy, Box 516, SE-751 20 Uppsala, Sweden}
\affiliation{Lyman Laboratory of Physics, Harvard University, Cambridge, Massachusetts 02138, USA}

\date{\today}

\begin{abstract}
The coexistence of charge density wave (CDW) and superconductivity in tantalum disulfide (2H-TaS$_{2}$) at ambient pressure, is boosted by applying hydrostatic pressures up to 30\,GPa, thereby inducing a typical dome-shaped superconducting phase. The ambient pressure CDW ground state which begins at $T_{CDW} \sim$ 76\,K, with critically small Fermi surfaces, was found to be fully suppressed at $P_{c} \sim$ 8.7\,GPa. Around $P_{c}$, we observe a superconducting dome with a maximum superconducting transition temperature $T_{c}$ = 9.1\,K. First-principles calculations of the electronic structure predict that, under ambient conditions, the undistorted structure is characterized by a phonon instability at finite momentum close to the experimental CDW wave vector. Upon compression, this instability is found to disappear, indicating the suppression of CDW order. The calculations reveal an electronic topological transition (ETT), which occurs before the suppression of the phonon instability, suggesting that the ETT alone is not directly causing the structural change in the system. The temperature dependence of the first vortex penetration field has been experimentally obtained by two independent methods and the corresponding lower critical field $H_{c1}$ was deduced. While a $d$ wave and single-gap BCS prediction cannot describe our $H_{c1}$ experiments, the temperature dependence of the $H_{c1}$ can be well described by a single-gap anisotropic $s$-wave order parameter.

\end{abstract}


\maketitle

Coexistence of superconductivity with competing physical phenomena such as magnetic or charge order has been of interest for the condensed matter community for a long time\cite{N0,N1,N3}. A commonly accepted argument says that for the materials exhibiting competing ground states, suppressing the magnetic or charge order helps to stabilize the superconducting (SC) phase. This is the case, for instance, in layered materials that are composed of two-dimensional (2D) building blocks, with periodic modulations of the charge carrier density, so-called charge density waves (CDWs)\cite{N44,N6,N8}. Classic examples are the members of the transition-metal dichalcogenide family (TMDs) MX$_{2}$, where M = Nb, Ti, Ta, Mo and X = S, Se. TMDs provide an ideal playground for studying semiconductors, metals, and superconductors in 2D using the same structural template\cite{N10,N11,N12}. For all known quasi-2D superconductors\cite{PG}, the origin and exact boundary of the electronic orderings and superconductivity are still attractive problems. At ambient pressure and without intercalation or chemical substitution, 2H-TaS$_{2}$, a prominent member of the vast family of TMDs, exhibits both superconductivity and a canonical CDW phase transition whose mechanisms remain controversial, even after decades of research\cite{Ta1,Ta2,Cava1}. Despite extensive studies, the current understanding of the microscopic origin of the SC mechanism and the CDW state is not complete. The SC transition temperature ($T_{c}$) increases while the CDW lock-in temperature falls down with chemical doping\cite{Cava1}, increasing  thickness of the sample\cite{ggg} and external pressure\cite{gg1,ZH,P1}. Several theoretical mechanisms behind the formation of CDW have been proposed\cite{1}. For TMDs, the following origins were extensively discussed: Fermi surface nesting\cite{2}, saddle points near Fermi surface\cite{3}, exciton-phonon\cite{4} or electron-phonon coupling\cite{5,6,7,8,9}. The most recent experimental evidence suggests that the latter plays a decisive role for CDW stabilization in Ta systems\cite{10,11}. It is thus of profound importance to understand the interplay between electronic and crystal structure in 2H-TaS$_{2}$. Additionally, there is no general consensus on the origin of SC pairing mechanisms in this material and further studies are necessary to elucidate this issue.

\begin{figure*}[!t]
\includegraphics[width=2\columnwidth]{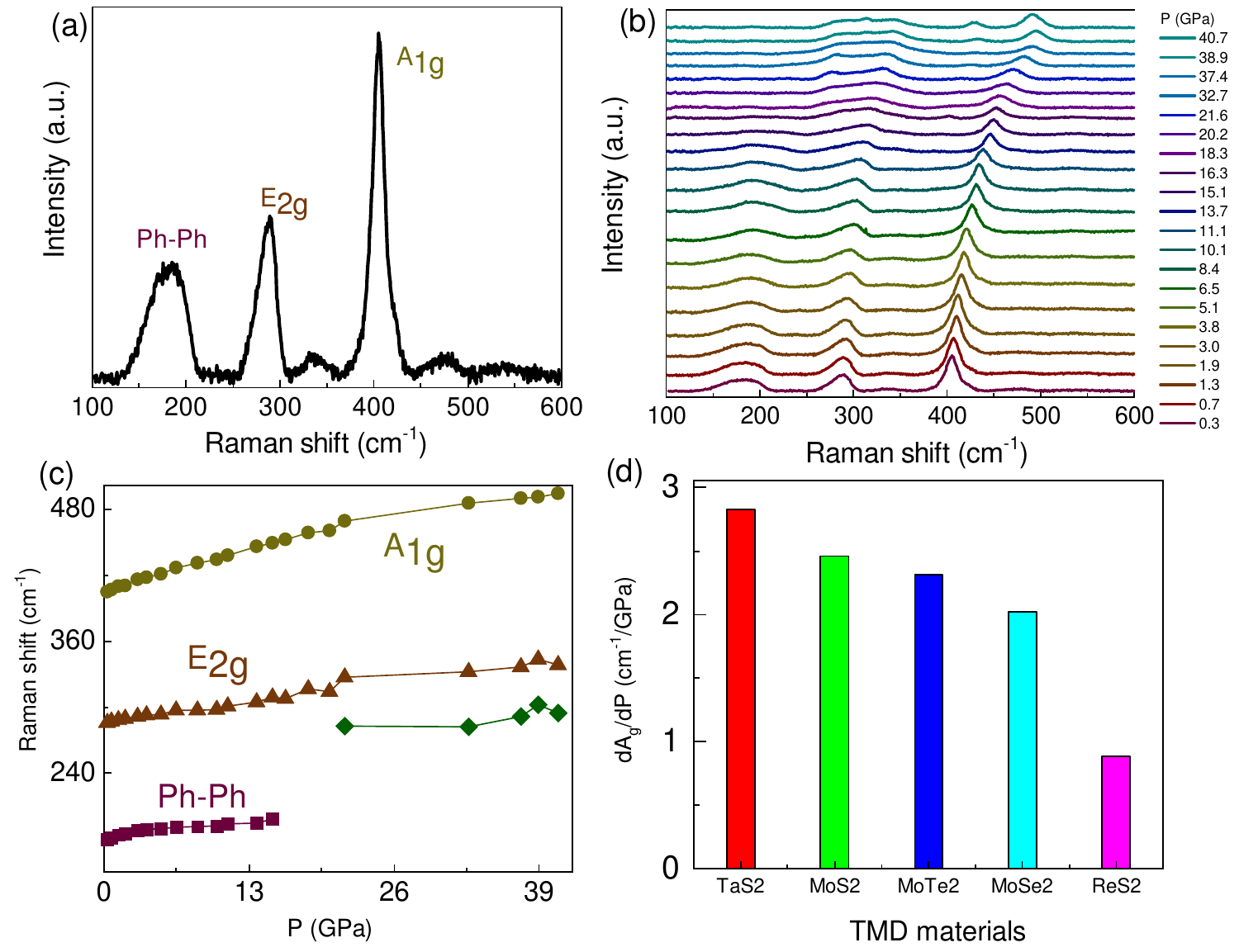}
\caption{(a) Raman scattering spectra of 2H-TaS$_{2}$ under room temperature and ambient pressure. (b) Raman of spectra under various hydrostatic pressures up to 40 GPa. (c) Pressure dependence of various vibrational modes in 2H-TaS$_{2}$. Ph-Ph refers to two-phonon mode. (d) Comparison of pressure coefficient (dA1g/dP) of the out-of-plane (A1g) Raman peaks of TaS$_{2}$, MoS$_{2}$, MoSe$_{2}$, MoTe$_{2}$ and ReS$_{2}$\cite{Nat}. One can see that the out-of-plane mode of TaS$_{2}$ presents the highest pressure coefficient among all selected materials.}
\label{fig:calc}
\end{figure*}

Whatever the proposed understanding of the relation between CDW and superconductivity is, it is important to determine the exact dependence of $T_{c}$ and the CDW phase with pressure\cite{Ta1}. Within this scope, through a combined complementary experimental techniques supplemented with theoretical calculations on 2H-TaS$_{2}$, we derive a previously not discussed pressure-temperature phase diagram. We explore external pressure as a tool to tune the phonon dispersions and thus the stability of the CDW phase. Pressure has long been recognized as a fundamental thermodynamic variable, and it is considered a very clean way to tune basic electronic and structural properties without changing the stoichiometry of a material\cite{M1,M2}. Our analysis shows that the temperature dependence of the lower critical fields, $H_{c1}(T)$, is inconsistent with a simple isotropic $s$-wave type of the order parameter but are rather in favor of the presence of an anisotropic $s$-wave. These observations clearly show that the SC energy gap in 2H-TaS$_{2}$ is nodeless.


\begin{figure*}[!t]
\includegraphics[width=1.8\columnwidth]{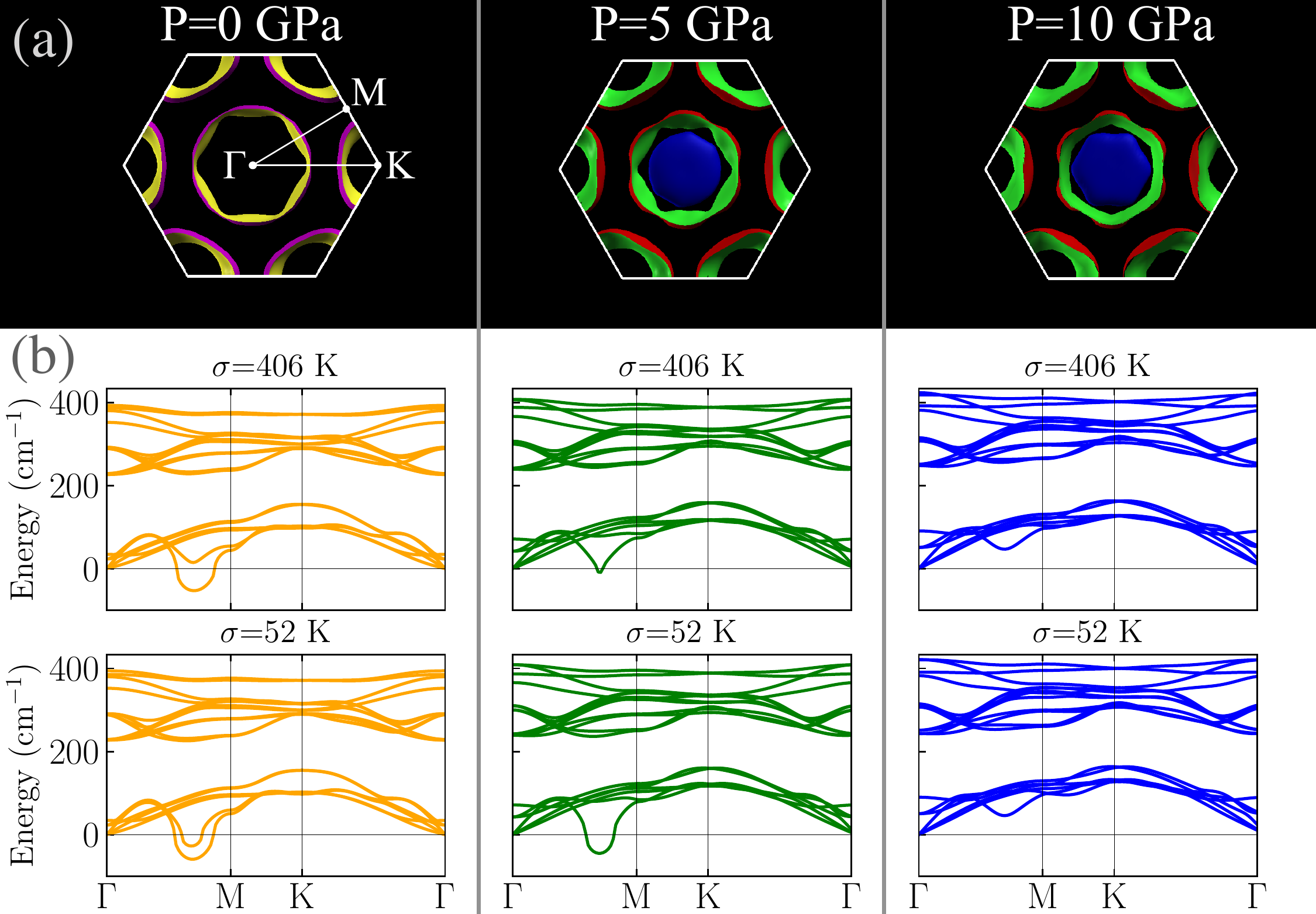}
\caption{Calculated Fermi surfaces (panel "a") and phonon dispersions (panel "b") for three different values of external pressure. The phonon dispersions were calculated for two different electronic temperatures, defined by the smearing parameter $\sigma$. The Fermi surface was plotted using XCrySDen software (see text for more details).}
\label{fig:calc}
\end{figure*}

\begin{figure}[tbp]
\includegraphics[width=20pc,clip]{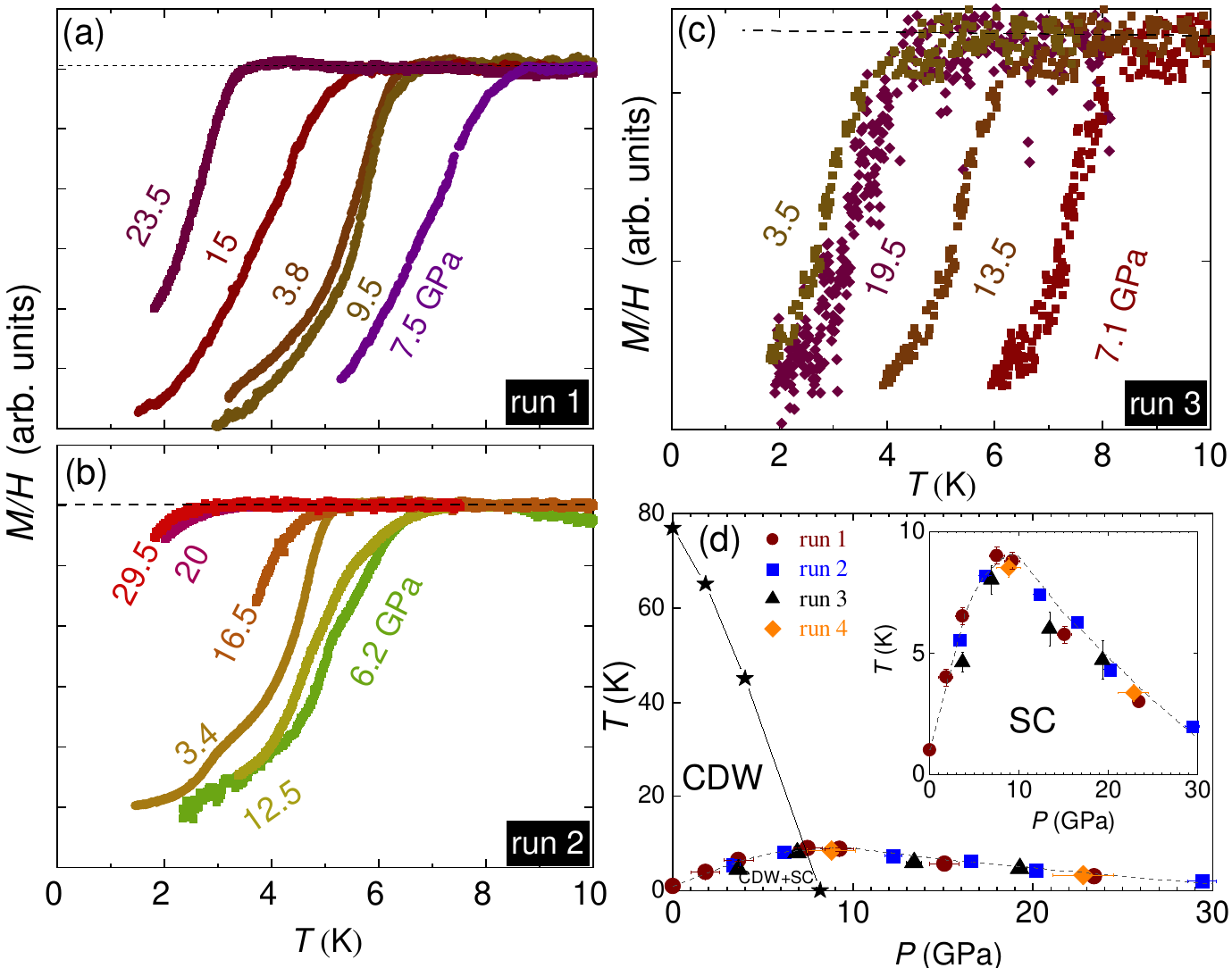}
\caption{The temperature dependence of the DC-susceptibility components of 2$H$-TaS$_{2}$ measured in dc field with an amplitude of 30\,Oe at elevated pressures. (a) $M$-$T$ curves at pressures between 3.8 and 23.5\,GPa in run 1. (b) $M$-$H$ curves at pressures between 6.2 and 29.5\,GPa in run 2. (c) $M$-$HT$ curves at pressures between 3.5 and 19.5\,GPa in run 3. The experimental data of run 4 is presented in Fig.\,S6\cite{17}. The data were collected upon warming in different the dc magnetic fields after cooling in a zero magnetic field. (d) The obtained pressure-temperature ($P$-$T$) phase diagram of 2$H$-TaS$_{2}$. Pressure dependence of the SC transition temperatures $T_{c}$ up to 30\,GPa. The values of $T_{c}$ were determined from the high-pressure resistivity and DC magnetic susceptibility\cite{Ta1,17}. The temperature dependence on the disappearance of the CDW as function of pressure, is shown as stars.}
\end{figure}

Details about the high-pressure measurements, crystal structure, and first-principles calculations can be found in the Supplemental Material\cite{17}. Raman response of 2H-TaS$_{2}$ at ambient pressure and room temperature is presented in Fig.\,1(a), where three regular phonons are observed: (i) a second-order peak due to two-phonon process at 180.3\,cm$^{-1}$ (ii) $E_{2g}$ - an in-plane vibrational mode at 288.1\,cm$^{-1}$ and $A_{1g}$ -out-of-plane mode at 405.4\,cm$^{-1}$, and these values are well agreed with the reported works\cite{Ra1,Th2}. Figure 1(b) shows the Raman spectra of 2H-TaS$_{2}$ under hydrostatic pressure up to 40\,GPa. The diamond background of each data point was subtracted by baseline fittings. By application of pressure, all Raman modes loose their intensities, get wider and show blue shift. Moreover, we could observe a splitting of the $E_{2g}$ peak which may be due to the pressure-induced structural phase transition in 2H-TaS$_{2}$. This mode is known to experience the discontinuity as a function of temperature, as one crosses $T_{CDW}$. The positions of all the three peaks against pressure are shown in Fig.\,1(c). Raman mode at 180.3 which represents two-phonon evolves till 15.1\,GPa, whereas other peaks exist till final pressure of 40\,GPa. For the case of $A_{1g}$ mode, it shows blue shift with deformation coefficient of 2.83 cm$^{-1}$/GPa. Pressure coefficient of $A_{1g}$ modes of similar TMDs are compared in a bar diagram [Fig.\,1(d)], and it can be clearly seen that 2H-TaS$_{2}$ is very sensitive to the hydrostatic pressure compared to other TMDs. As ReS$_{2}$ is vibrationally decoupled, its Raman spectrum is less sensitive to pressure. On the other hand, as $A_{1g}$ mode is vibrationally coupled with $E_{2g}$ mode in the case of TaS$_{2}$, pressure coefficient of $A_{1g}$ mode of 2H-TaS$_{2}$ is higher than in any other TMD materials. As the pressure increases, pressure coefficient of $A_{1g}$ is reduced above 20\,GPa at which structural transition may be expected. For instance, pressure would significantly reduce interlayer distance, so adjacent layers will be coupled and overlap of electron wavefunctions is stronger in TaS$_{2}$ than in MoS$_{2}$ and band structure transformation may happen above 20\,GPa.


\begin{figure}[tbp]
\includegraphics[width=20pc,clip]{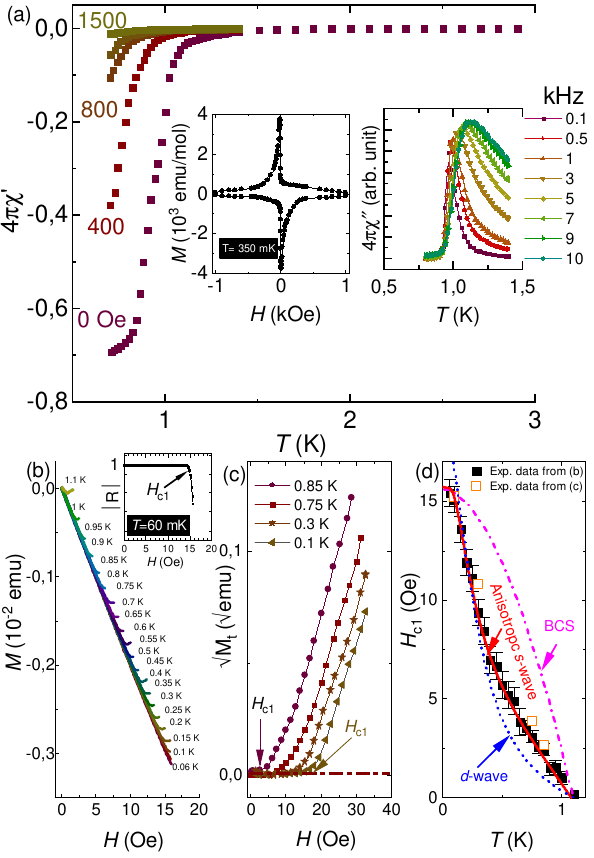}
\caption{(a) The temperature dependence of the complex AC-susceptibility components $4\pi\chi{'}_{v}$ measured in an AC field with an amplitude of 5\,Oe and a frequency of 1\,kHz. Data were collected upon warming in different DC magnetic fields after cooling in a zero magnetic field. The insets illustrate the isothermal magnetization $M$ vs. $H$ loops measured at 350\,K up to 1000\,Oe applied along the $c$-axis and the imaginary part of AC at various frequency ($\nu _{m}$). (b) The SC initial part of the magnetization curves measured at various temperatures down to 60\,mK. The inset depicts an example used to determine the $H_{c1}$ value using the regression factor $R$, at $T$ = 60\,mK\cite{17}. (c) The field dependence of $\sqrt{M_{t}}$ at various temperatures. The arrows indicate $H_{c1}$ values that estimated by extrapolating the linear fit of $\sqrt{M_{t}}$ to 0. (d) Phase diagram of $H_{c1}$ for the field applied parallel to the $c$ axis. $H_{c1}$ has been estimated by two different methods from the extrapolation of $\sqrt{M_{t}}$ to 0 (open symbols) and from detecting the transition from a Meissner-like linear (closed symbols). The solid red line is the fitting curves using anisotropic $s$-wave approach. The dotted and dashed lines represent the $d$-wave and a single-gap BCS approach, respectively.}
\end{figure}

In order to get a physical insight into the suppression of the CDW phase under pressure, we have investigated the electronic structure by means of \textit{ab initio} theory. Recently, first-principles calculations show that the electron-phonon interactions depend on both the amount of applied strain and the direction in 2H-TaSe$_2$\cite{Thr}. In addition, a sudden change in $E_{2g}$ mode in 2H-TaSe$_2$ is observed\cite{11}. The main results are shown in Fig.\,2. According to the results obtained for an undistorted 2H-TaS$_2$, the material undergoes a pressure-induced electronic topological (so-called Lifshitz\cite{Th1}) transition. An additional hole pocket around the $\Gamma$ point emerges, as shown in blue on Fig.\,2(a). This transition happens below 2.5\,GPa, and upon further compression, at least up to 15\,GPa, the Fermi surface topology is intact, while its shape becomes slightly modified. The pressure evolution of the calculated phonon spectra is shown in Fig.\,2(b). At the equilibrium, and under small applied pressures, there is a phonon instability along the $\Gamma-M$ direction at the wavevector close to experimental $q_{\text{CDW}}$. Upon compression, the instability is suppressed somewhere between 5 and 10\,GPa, indicating the suppression of the CDW order.

Interestingly, the instability disappears \textit{after} the ETT, which indicates that the Fermi surface nesting itself is not the only driving force of CDW order, which is in line with other, more recent studies on 2H-TaS$_2$\cite{10}. The results of our calculations for ambient pressure are in agreement with Ref.\cite{Th2}. These types of calculations are not able to properly capture the CDW transition temperature, but show a correct qualitative behaviour. As the electronic temperature increases, the phonon instability becomes less pronounced, but persists up the temperatures well above experimental CDW ordering temperature. The authors of Ref.\cite{Th2} attribute this to the presence of a short-range CDW state. We would like to note, however, that the employed treatment of the temperature effects is potentially oversimplified and does not capture many phenomena. The main reason is an incomplete description of the electronic correlations within DFT. Moreover, certain crystal structures are known to be stabilized due to anharmonic effects\cite{Th3}. An explicit account of the electron-phonon interaction which is expected to be quite anisotropic\cite{10} might induce strong modifications of both electronic and phononic spectra.

The effects of 8.7\,GPa illustrate a suppression of the CDW state and enhanced the $T_{c}$ with a very sharp drop of the resistivity up to 9.1\,K\cite{17}. Figure 3(a-c) shows the temperature-dependent magnetic $M$(T) of the 2$H$-TaS$_{2}$ at pressures from 0 to 30\,GPa in three runs. The dome-like evolution of $T_{c}$ was constructed based on the observed pressure-dependent magnetization data shown in Fig.\,3(d), which explicitly shows the gradual suppression of the CDW phase. As displayed in Fig.\,3(a-c), it is clear that $T_{c}$ increased up to a pressure of 8.5\,GPa, where it exhibits a maximum, then immediately begins to turn down. This kind of dome-shaped curve is one of the hallmarks of high temperature superconductors, but many mysteries around these types of domes remain to be explained\cite{W1}. Deriving a solid picture of the origin of the SC dome constitutes a major challenge. Applying external pressure to the system simply modifies the interatomic spacing, wavefunction overlap and electronic structure, as well as the balance between kinetic energy and Coulomb interaction among the electrons. The SC state certainly depends on these parameters, which is determined by both the pressure and the existence of the CDW state. Since both pressure and the CDW state heavily influence $T_{c}$, the competition between the two might be the cause of the SC dome.

At ambient pressure, 2$H$-TaS$_{2}$ exhibits a prominent CDW anomaly at 76\,K. While superconductivity is well distinguished by the resistivity and specific heat measurements\cite{17}, we further confirmed the bulk superconductivity by performing low-temperature AC susceptibility, $\chi{'}$,  measurements as illustrated in Fig.\,4(a). $T_{c}$ of 1.2\,K has been extracted from the bifurcation point between $\chi{'}_{v}$ and $\chi{''}_{v}$. One can clearly see that the maximum of the temperature dependence of the imaginary part of AC susceptibility, see Fig.\,4(a) (right inset), shifts to higher temperatures upon increasing the frequency which we attribute to the motion of vortices. The lower critical field, $H_{c1}$, i.e. the thermodynamic field at which the presence of vortices into the sample becomes energetically favorable, is a very useful parameter providing key information regarding bulk thermodynamic properties. Therefore, a reliable determination of the lower critical field, $H_{c1}$, from magnetization measurements has been determined. The most popular approach of determining the $H_{c1}$, is the point of deviation from a linear M(H) response, compared with the values obtained from the onset of the trapped magnetic moment $[$($M_{t}$ Fig.\,4(c)$]$, (see\cite{17} for more details). We have confirmed the absence of the surface barriers in our case from the very symmetric DC magnetization hysteresis curves M(H) at 350\,mK $[$Fig.\,4(a) (left inset)$]$. The experimental values of $H_{c1}$ were corrected by accounting for the demagnetization effects. Indeed, the deflection of field lines around the sample leads to a more pronounced Meissner slope given by $M/H_{a} = -1/(1-N)$, where $N$ is the demagnetization factor and is found to be $\approx0.97$. Taking into account these effects, the absolute value of $H _{c1}$ can be estimated by using the relation proposed by Brandt\cite{Brandt}. The most intriguing feature in Fig.\,4(d) is the upward trend with negative curvature over the entire temperature range, similar features are reported in\cite{CR,Yadav}.

With the above understanding of the nature of superconductivity in 2$H$-Ta$S_2$, We now turn to study its gap symmetry and structure of the SC order parameter, which can be used to reveal the pairing mechanism. The obtained experimental temperature dependence of $H _{c1}$ shown in Fig.\,4(d) was analyzed using the phenomenological $\alpha$-model. This model generalizes the temperature dependence of gap to allow $\alpha=2 \Delta(0)/T_c > 3.53$ (i.e. $\alpha$ values higher than the BCS value). The temperature dependence of each energy gap for this model can be approximated as:~\cite{BCS,Carrington}  

$\Delta _{i}(T) = \Delta _{i}(0) {\tanh[1.82(1.018(\frac{T_{ci}}{T}-1))^{0.51}]}$, 

where $\Delta(0)$ is the maximum gap value at $T$ = 0. We adjust the temperature dependence of $H_{c1}$, which relates to the normalized superfluid
density as $\tilde{\rho _{s}}(T)$=$H _{c1}(T)$/$H _{c1}$(0), by using the following expression:
\begin{equation}
\label{eq2} \frac{H_{c1}(T)}{H_{c1}(0)} = 1+\frac{1}{\pi}\int^{2 \pi}_0{ 2\int_{\Delta(T,\phi)}^{\infty}{\frac{\partial f}{\partial E} \frac{E dE d\phi}{\sqrt{E^2-\Delta^2(T,\phi)}}}},
\end{equation}
where $f$ is the Fermi function $ [ \exp( \beta E + 1 )]^{-1}$, $\varphi$ is the angle along the Fermi surface, $\beta$ = ($k_\textup{B}T)^{-1}$. The energy of the quasiparticles is given by $E$ = $[\epsilon^{2} + \Delta^{2}(t)]^{0.5}$, with $\epsilon$ being the energy of the normal electrons relative to the Fermi {level}, and where $\Delta(T,\phi)$ is the order parameter as function of temperature and angle. We used for the $s$-wave, $d$-wave the following expressions  $\Delta(T,\phi)=\Delta(T)$ and $\Delta(T,\phi)=\Delta(T) \cos(2 \theta)$, respectively. The main features from the corrected $H _{c1}$ values in Fig.\,4(d) can be described in the following way: (i) As a first step we compare our data to the single band $s$-wave and we find a systematic deviation at high temperature data, (ii) More obvious deviations exist in the case of $d$-wave approach\cite{Carrington}. This clearly indicates that the gap structure of our system is more likely to be nodeless $s$-wave, (iii) Then, anisotropic $s$-wave is further introduced to fit the experimental data. For the anisotropic $s$-wave, the fitting  with the magnitude of the gap $\Delta _0$ = 1.21\,meV with an anisotropy parameter $\approx$1.01. As can be seen the anisotropic $s$-wave order parameter presents a good description to the data. We hence conclude that in Ta$S_2$ the exotic SC gap structure is related to the Ta tubular sheets and that, even if the charge density wave is perturbing those sheets in TaS$_2$, this CDW does not affect the SC gap structure.

The temperature-pressure phase diagram of Ta$S_2$ is demonstrated here to have a dome-like SC phase with a maximum SC transition temperature $T_{c}$ = 9.1\,K. By employing \textit{ab initio} electronic structure theory, we were able to investigate the temperature and pressure dependence of the phonon spectrum. It is shown that, at ambient conditions, there is a phonon instability at the propagation vector close to the $q_{\text{CDW}}$ wavevector. Furthermore, the temperature dependence measurements of the critical field are consistent with single gap anisotropic $s$-wave superconductivity.

YOK and MAH acknowledge  the  financial  support  from  the Swedish  Research  Council  (VR)  under  the  project  No.2019-03569 and 2018-05393. We acknowledge Rajeev Ahuja and Goran Karapetrov for helpful discussions.


\begin{thebibliography}{100}

\bibitem{N0} T. Yokoya, T. Kiss, A. Chainani, S. Shin, M. Nohara and H. Takagi, Science \textbf{294}, 2518-2520 (2001).
\bibitem{N1} M. B. Maple, Appl. Phys. \textbf{9}, 179 (1976).
\bibitem{N3} Paul C. Canfield, Peter L. Gammel, and David J. Bishop, Physics Today \textbf{51} (10), 40 (1998).

\bibitem{N44} M. D. Johannes and I. I. Mazin, Phys. Rev. B \textbf{77}, 165135 (2008).
\bibitem{N6} J. J. Hamlin, D. A. Zocco, T. A. Sayles, M. B. Maple, J. H. Chu, and I. R. Fisher, Phys. Rev. Lett. \textbf{102}, 177002 (2009).
\bibitem{N8} D. A. Zocco, J. J. Hamlin, K. Grube, J. H. Chu, H.H.Kuo, I. R. Fisher, and M. B. Maple, Phys. Rev. B \textbf{91}, 205114 (2015).
\bibitem{N10} B. Sipos, A. F. Kusmartseva, A. Akrap, H. Berger, L. Foro, and E. Tutis, Nat. Mater. \textbf{7}, 960 (2008).
\bibitem{N11} Q. H. Wang, K. Kalantar-Zadeh, A. Kis, J. N. Coleman, and M. S. Strano, Nat. Nanotechnol. \textbf{7}, 699 (2012).
\bibitem{N12} A. K. Geim and I. V. Grigorieva, Nature (London) \textbf{499}, 419 (2013).
\bibitem{PG} A. A. Kordyuk, Low Temp. Phys. / Fiz. Nizk. Temp. \textbf{41}, 417 (2015).

\bibitem{Ta1} M. Abdel-Hafiez, X.-M. Zhao, A. A. Kordyuk, Y.-W. Fang, B. Pan, Z. He, C.-G. Duan, J. Zhao, and  X.-J. Chen, Sci. Rep. \textbf{6}, 31824 (2016)
\bibitem{Ta2} D. C. Freitas et al., Phys Rev. B \textbf{93}, 184512 (2016).
\bibitem{Cava1} K. E. Wagner, E. Morosan, Y. S. Hor, J. Tao, Y. Zhu, T. Sanders, T. M. McQueen, H. W. Zandbergen, A. J. Williams, D. V. West, and R. J. Cava, Phys. Rev. B \textbf{78}, 104520 (2008).
\bibitem{ggg} Y. Yu, F. Yang, X. F. Lu, Y. J. Yan, Y.-H. Cho, L. Ma, X. Niu, S. Kim, Y.-W. Son, D. Feng, S. Li, S.-W. Cheong, X. H. Chen and Y. Zhang, Nat. Nano. \textbf{10}, 270 (2015).
\bibitem{gg1} C. Berthier, P. Molinie, and D. Jerome, Solid State Commun. \textbf{18}, 1393 (1976).
\bibitem{ZH} Z.-H. Chi, X.-M. Zhao, H. Zhang, A. F. Goncharov, S. S. Lobanov, T. Kagayama, M. Sakata, and X.-J. Chen,  Phys. Rev. Lett. \textbf{113}, 036802 (2014).
\bibitem{P1} B. Sipos, A. F. Kusmartseva, A. Akrap, H. Berger, L. Forro, and E. Tutis, Nat. Mat. \textbf{7}, 960 (2008).

\bibitem{1} X. Zhu, J. Guo, J. Zhang, and E. W. Plummer, Advances in Physics: X \textbf{2}, 622 (2017).
\bibitem{2} J. A. Wilson, F. J. Di Salvo, and S. Mahajan, Phys. Rev. Lett. \textbf{32}, 882 (1974).
\bibitem{3}T. M. Rice and G. K. Scott, Phys. Rev. Lett. \textbf{35}, 120 (1975).
\bibitem{4} J. van Wezel, P. Nahai-Williamson, and S. S. Saxena, Phys. Rev. B \textbf{81}, 165109 (2010).
\bibitem{5} A. H. Castro Neto, Phys. Rev. Lett. \textbf{86}, 4382 (2001).
\bibitem{6} F. Weber, S. Rosenkranz, J.-P. Castellan, R. Osborn, R. Hott, R. Heid, K.-P. Bohnen, T. Egami, A. H. Said, and D. Reznik, Phys. Rev. Lett. \textbf{107}, 107403 (2011).
\bibitem{7} M. D. Johannes, I. I. Mazin, and C. A. Howells, Phys. Rev. B \textbf{73}, 205102 (2006).
\bibitem{8} M. D. Johannes and I. I. Mazin, Phys. Rev. B \textbf{77}, 165135 (2008).
\bibitem{9} L. P. Gor'kov, Phys. Rev. B \textbf{85}, 165142 (2012).
\bibitem{10} K. Wijayaratne, J. Zhao, C. Malliakas, D. Young Chung, M. G. Kanatzidis, and U. Chatterjee, J. Mater. Chem. C \textbf{5}, 11310 (2017).
\bibitem{11} H. M. Hill, S. Chowdhury, J. R. Simpson, A. F. Rigosi, D. B. Newell, H. Berger, F. Tavazza, and A. R. Hight Walker, Phys. Rev. B \textbf{99}, 174110 (2019).
\bibitem{Nat} S. Tongay, H. Sahin, C. Ko, et al. Nat Commun \textbf{5}, 3252 (2014).

\bibitem{M1} M. Abdel-Hafiez, M Mito, K Shibayama, S Takagi, M Ishizuka, AN Vasiliev, C Krellner, H. Mao, Phys. Rev. B \textbf{98}, 094504 (2018).
\bibitem{M2} M. Abdel-Hafiez, Y. Zhao, Z. Huang, C. Cho, C. Wong, A. Hassen, M. Ohkuma, Y. Fang, B. Pan, Z. Ren, A. Sadakov, A. Usoltsev, V. Pudalov, M. Mito, R. Lortz, C Krellner, W Yang, Phys. Rev. B \textbf{97}, 134508 (2018).

\bibitem{17} Supplementary Material is available for the experimental and calculation details and the supporting results.

\bibitem{Ra1} R. Grasset, Y. Gallais, A. Sacuto, M. Cazayous, S. Manas-Valero, E. Coronado, and M.-A. Measson, Phys. Rev. Lett. \textbf{122}, 127001 (2019).

\bibitem{Th2} J. Joshi, H. M. Hill, S. Chowdhury, C. D. Malliakas, F. Tavazza, U. Chatterjee, A. R. HightWalker, and P.M. Vora, Phys. Rev. B \textbf{99}, 245144 (2019).

\bibitem{Th1} I. M. Lifshitz, Sov. Phys. JETP \textbf{11}, 1130 (1960).

\bibitem{Thr} S. Chowdhury, J. R. Simpson, T. L. Einstein, and A. R. H. Walker, Phys. Rev. Mat. \textbf{3}, 084004 (2019).

\bibitem{Th3} P. Souvatzis, O. Eriksson, M. I. Katsnelson, and S. P. Rudin, Phys. Rev. Lett. \textbf{100}, 095901 (2008).

\bibitem{W1} J. Lu, O. Zheliuk, Q. Chen, I. Leermakers, N. E. Hussey, U. Zeitler, J. Ye. PNAS 201716781 (2018) DOI: 10.1073/pnas.1716781115

\bibitem{Brandt} E. H. Brandt, Phys. Rev. B \textbf{60}, 11939 (1999).

\bibitem{CR} C. Ren, Z.-S. Wang, H.-Q. Luo, H. Yang, L. Shan, H.-H. Wen, Phys. Rev. Lett. \textbf{101 } 257006, (2008).

\bibitem{Yadav} C. S. Yadav, and P. L. Paulose, New Journal of Physics \textbf{11}, 103046 (2009).

\bibitem{BCS} A. Carrington and F. Manzano, Physica C \textbf{385}, 205 (2003).

\bibitem{Carrington} A. Carrington, and F. Manzano, Physica C \textbf{385}, 205 (2003).

\end{thebibliography}
\end{document}